\documentclass[traditabstract]{aa}
\usepackage{graphicx}
\usepackage{natbib,twoopt}
\usepackage{amssymb}
\usepackage{amsfonts}
\usepackage{amsbsy}
\usepackage{amsmath}
\usepackage{color}

\newcommand{\be}{\begin{equation}}
\newcommand{\ee}{\end{equation}}

\newcommand{\etal}{et al.\ }

\newcommand{\Sigmab}{\boldsymbol{\Sigma}}
\newcommand{\pb}{\boldsymbol{p}}
\newcommand{\phb}{\boldsymbol{\widehat{p}}}

\newcommand{\Gb}{\boldsymbol{G}}
\newcommand{\Ib}{\boldsymbol{I}}

\newcommand{\phih}{\widehat{\phi}}

\newcommand{\rhoh}{\widehat{\rho}}
\newcommand{\thetah}{\widehat{\theta}}
\newcommand{\Phih}{\widehat{\Phi}}

\newcommand{\Thetah}{\widehat{\Theta}}
\begin{document}
\title{The bivariate K-band-submillimetre luminosity functions of the local HRS galaxy sample}
\titlerunning{submm local luminosity function}

   \author{P. Andreani \inst{1} 
 	L.~Spinoglio\inst{2}, A.~Boselli\inst{3}, L.~Ciesla\inst{3}, L.~Cortese\inst{4}, R.~Vio\inst{5}, M.~Baes\inst{6}, G.J.~Bendo\inst{7}, I.~De~Looze\inst{6}}
	\institute{European Southern Observatory, Karl-Schwarzschild-Stra\ss e 2, 85748 Garching, Germany, \email{pandrean@eso.org}
    \and
         INAF - Istituto di Astrofisica e di Planetologia Spaziale, via del Fosso del Cavaliere, Roma, Italy 
     \and
         Laboratoire d'Astrophysique de Marseille - LAM, Universit\'e d'Aix-Marseille \& CNRS, UMR7326, 38 rue F. Joliot-Curie, 13388 Marseille Cedex 13, France 
\and
        Centre for Astrophysics \& Supercomputing, Swinburne University of Technology, Mail H30 P.O. Box 218 Hawthorn, 3122 Victoria, Australia
\and        
      Chip Computers Consulting s.r.l., Viale Don L.~Sturzo 82, S.Liberale di Marcon, 30020 Venice, Italy
        \and
        Sterrenkundig Observatorium, Universiteit Gent Krijgslaan 281 S9, B-9000 Gent, Belgium 
        \and
        UK ALMA Regional Centre Node, Jodrell Bank Centre for Astrophysics, School of Physics and Astronomy, University of Manchester, Oxford Road, Manchester M13 9PL, UK}
\authorrunning{Andreani, Spinoglio \etal}

\date{ }
\abstract{We study the relationship between the {\it K-band} and the sub-millimetre (submm) emissions of nearby galaxies by computing the bivariate {\it K-band}-submm luminosity function (BLF) of the {\textit{Herschel}}~\thanks{\textit{Herschel is an ESA space observatory with science instruments provided by European-led Principal Investigator consortia and with important participation from NASA.}} Reference Survey (HRS), a volume-limited sample observed in submm with {\textit{Herschel}}/SPIRE.
We derive the BLF from the {\it K-band} and submm cumulative distributions using a {{\it copula}} method.
Using the BLF allows us to derive the relationship between the luminosities on more solid statistical ground. The analysis shows that over the whole HRS sample, no statistically meaningful conclusion can be derived for any relationship between the {\it K-band} and the submm  luminosity. However, a very tight relationship between these luminosities is highlighted, by restricting our analysis to late-type galaxies. The fuminosity function of late-type galaxies computed in the {\it K-band} and in the submm are dependent and the dependence is caused by the link, between the stellar mass and the cold dust mass, which has been already observed.
}
\keywords{Galaxies: luminosity function -- Galaxies: nearby galaxies --
Galaxies: physical process -- Galaxies: star formation --
Infrared: Galaxies}
\maketitle

\date{Received .............; accepted ................}

\section{Introduction}
One of the most used tools in astronomy for probing the distribution of luminous matter in the Universe over cosmic time is the luminosity function (LF), which is the probability distribution function (PDF) of galaxy luminosities, the relative number of galaxies of different luminosities in a representative volume of the Universe \citep[see e.g.][for a recent review of the subject]{Johnston11}.
To accurately constrain the LF, a complete sample is required, a sample of objects selected at given frequencies with well defined selection criteria for which biases
are understood well and accounted for. Biases are usually introduced when selecting a sample, if the selection criteria favour a particular class or population of objects.
By incompleteness we mean a lack of a uniform coverage in redshift, in flux or in any other physical quantity \citep[][]{SmithR12}.
 
To determine the LF, one starts from a sample selection at a given wavelength band and counts the number of galaxies in that band. If galaxies are selected in a particular band, but the number counts are carried out at a different wavelength, one needs to control possible selection effects in both bands before deriving any meaningful properties of the LF and its underlying sample \citep[i.e.][]{Bose11}.
In this case one can construct the  bivariate LF (BLF), which explores the dependency of the LFs at different wavelengths on each other, and explore the bivariate properties of some physical parameters of the sources \citep{Johnston11}.

In a wider more general context, a bivariate distribution is often used to investigate the relationships between variables of a given sample. Quite often the bivariate distribution is obtained by ad hoc methods or heuristically, but analytic models are then required to interpret the results \citep{Cholo85,Chap+03,Schaf07,CrossDriver02,Bal+06,Driver+06}. 
Inconsistent results for the relationship between two quantities are often obtained from regression analyses only because of the {\it a priori} assumptions on the
dependency between variables \citep{LaFranca+95}. The difficulty in assigning an {\it a-priori} distribution to a variable often results in a very poor fit from which no statistically meaningful conclusion can be derived \citep{Bal+06}.
  
However, constructing a BLF from experimental  data is not a trivial task. Mathematically, this problem implies reconstructing a bivariate PDF satisfying prescribed marginals (i.e. cumulative distribution functions).
If the distribution is not bivariate Gaussian, this operation is not straightforward. In fact, an infinite number of bivariate distributions exists with the same marginals, which can be constructed once the dependence structure is specified. The goal is, therefore, to find a way to disentangle the marginal distribution and the dependence structure. Here we adopt an approach based on the 
PDF of each random variable and the so-called {\rm copula}, a function that provides the dependence structures between them. We show in detail how it is possible to reconstruct a BLF from a multi-wavelength dataset.

In the past the BLF has been computed at optical/X-ray wavelengths by La Franca et al. (1995) 
and for the SDSS sample by Cross \& Driver (2002), Ball et al. (2006), and Driver et al. (2006). 
All these approaches were based on a maximum likelihood (ML) fit to a bivariate PDF, or its non-parametric version (the stepwise ML fit SWML introduced by Efstathiou et al. 1988). 
However, until recently there has not been any general method of constructing a bivariate distribution function with pre-defined marginal distributions and correlation coefficient. Our approach is fundamentally different from previous works dealing with the bivariate LF. We follow Takeuchi (2010), 
who introduced a {\rm copula} method to estimate the bivariate LF in the infrared, but we use a different implementation from these authors, mainly in the computation of the correlation coefficient.


We compute the bivariate {\it K-band}-submm luminosity function of the {\textit{Herschel}} Reference Survey (HRS) sample \citep{Bose+10a}. This sample has been extracted from the 2MASS survey and observed in the submm by the SPIRE instrument on board of the {\textit{Herschel}} satellite. 
We make use of an analytical function of the {\it K-band} and submm LFs, and we use it as the marginal PDF. As explained in the text, the choice we made is dictated by the smallness of the sample and by the presence of upper limits among the SPIRE observations.

Since our sample is inhomogeneous and relatively small, the direct derivation of the submm LF cannot be pursued \citep[as in e.g.][]{Dye+10, Dunne+11, Davies+10}, while we can exploit the knowledge of the {\it K-band} LF, which on the contrary is well established \citep{Cole+01,Kochanek+01}.
In addition using the BLF as a statistical tool in the presence of upper limits provides results that lie on more solid statistical ground than any other simpler tool,
i.e. a linear regression test. 

The paper is organised as follows. The methodology for computing the luminosity functions is described in section~\ref{tools}.
The sample selection is briefly described in section~\ref{sample} with the results shown in section~\ref{results} and a discussion in section ~\ref{discussion}.

\section{Mathematical tools}\label{tools}

\subsection{Definition of the luminosity functions}

The LF, $\phi(L)$, is defined as the number density of galaxies whose luminosity lies within the interval $[L, L+dL]$ 

\begin{equation}
\phi (L) = \frac{ dn}{d L}.
\end{equation}

\noindent
This definition of the LF implies that the integral over the luminosity provides the number of objects per unit volume, $n$. For the rest of the paper we use its normalised version, $\int \phi (L) d L =1$, which represents a PDF.

\subsection{Estimation of a bivariate luminosity function using a semi-parametric approach}\label{blf}

Given a set of $N$ {\it observed} quantities $\{ x_i \}_{i=1}^N$ and $\{ y_i \}_{i=1}^N$, determining of the bivariate PDF $\psi(x,y)$, such that $\psi(x,y) dx dy$
is the probability that a random variable (in this case the luminosity) takes values in the range $[x, x+dx]$ and $[y, y+dy]$, is not a trivial problem.
The extension of the classical estimators, such as the histogram to the two-dimensional case is not very productive since, to obtain satisfactory results, a much greater number of data is needed than in the one-dimensional case. 
As shown in the following, a potentially more useful alternative is an approach based on {\rm copulas} \citep[see][for the mathematical definition of {\rm copula}]{Schmidt07}.

We define the quantities $u_x=\Phi(x)$ and $u_y=\Theta(y)$, with
\begin{align}
\Phi(x) & = \int_{x'_{{\rm min}}}^x \phi(x') dx', \label{eq:Phi} \\
\Theta(y) & = \int_{y'_{{\rm min}}}^y \theta(y') dy', \label{eq:Theta}
\end{align}
the cumulative distribution function (CDF) of the PDFs $\phi(x)$ and $\theta(y)$ (hereafter called {\it marginals}), which are distributed according to a uniform distribution that takes values in the range $[0, 1]$. If $G^{-1}(u_z)$ is 
the inverse function of the standard Gaussian CDF
$G(z)$, the quantities $z_x$ and $z_y$:
\begin{align}
z_x & = G^{-1}(u_x), \label{eq:z1} \\
z_y & = G^{-1}(u_y), \label{eq:z2}
\end{align}
are distributed according to a standard Gaussian PDF, $g(z)$; i.e., they are Gaussian variables. In other words, by means of Eqs.~(\ref{eq:Phi})-(\ref{eq:z2}) the random variables $x$ and $y$ are  {\it Gaussianised}.
It is assumed that  the joint PDF $g_{\Sigmab}(z_x,z_y)$ of 
$z_x$ and $z_y$ is the bivariate Gaussian PDF with covariance matrix $\Sigmab$ given by
\begin{equation} \label{eq:Sigma}
\Sigmab=\left(
\begin{array}{cc}
1 & \rho \\
 \rho & 1
\end{array}
\right),
\end{equation}
where $\rho$ is the linear correlation coefficient of the two random variables $z_x$ and $z_y$. Such an approach is similar to what is proposed by Takeuchi et al. (2010), the only difference being the way the coefficient $\rho$ is estimated (see below).

The {\rm copula}  $C_{\Sigmab}(u_x, u_y )$ of $g_{\Sigmab}(z_x,z_y)$ is defined from the equation \citep[i.e.][]{Schmidt07}:

\begin{equation} 
\psi \left(x, y\right) = c(u_x, u_y) \phi\left(x\right) \theta\left(y\right), \label{eq:pdf}
\end{equation} 
where
$x=\Phi^{-1}(u_x)$ and $y=\Theta^{-1}(u_y)$ and 
\begin{equation} \label{eq:c}
c_{\Sigmab}(u_x, u_y) = \frac{\partial^2 C_{\Sigmab}(u_x, u_y)}{\partial u_x \partial u_y}.
\end{equation}
We recall that a $d$-dimensional {\rm copula} $C: [0, 1]^d \rightarrow [0, 1]$ is a CDF with uniform marginals. 
{\rm Copulas} are used to describe the dependence between random variables, and their main use is to disentangle marginals and the dependence structure.
The importance of this function lies in the fact that it describes the dependence structure between the random variables separated by the corresponding marginals.
In particular, with the Gaussian {\it copula} the dependence structure is parametrised by a single parameter, the correlation coefficient.

It is possible to see that
\begin{equation} \label{eq:C}
C_{\Sigmab}(u_x, u_y ) = G_{\Sigmab} \left( g^{-1}(u_x), g^{-1}(u_y) \right),
\end{equation}
and from Eq.~(\ref{eq:c})
\begin{equation} \label{eq:cg}
c_{\Sigmab} (u_x, u_y) = \frac{1}{| \Sigmab|} \exp{ \left\{ - \frac{1}{2} \left[ \Gb^{-T} (\Sigmab^{-1} - \Ib) \Gb^{-1}\right] \right\}}, 
\end{equation}
with $\Gb^{-1} \equiv \left( G^{-1}(u_x), G^{-1}(u_y) \right)^T$, where $\Gb^{-T}$ is the transpose of $\Gb^{-1}$, $\Ib$ the identity matrix, and  $|\Sigmab|$ the determinant of $\Sigmab$.
In summary, to obtain a full description of the two variables together two ingredients are needed: the marginals and the type of interrelation.


Using the above results, a procedure for estimating the bivariate PDF $\psi(x,y)$ in the presence of possible left-censored data (upper limits) is the following.
\begin{enumerate}
\item Estimation of the marginals $\phih(x)$ and $\thetah(y)$ by means of a ML fit to the data $\{ x_i \}_{i=1}^{N_{x_o}}$, $\{ x_j \}_{i=1}^{N_{x_c}}$, $\{ y_k \}_{k=1}^{N_{y_o}}$, and $\{ y_l \}_{l=1}^{N_{y_c}}$ 
of given parametric PDFs $\phi(x | \pb_{\phi})$ and $\theta(y | \pb_{\theta})$.
The estimate $\phb$ of the parameters $\pb$ is given by
\begin{align}
\phb_{\phi} & = \underset{ \pb_{\phi} }{\arg\max}  \left[ \prod_i \phi(x_i | \pb_{\phi})  \prod_j \int_{x_{\rm min}} ^{x_j} \phi(x | \pb_{\phi}) dx \right], \label{eq:ML1}\\
\phb_{\theta} & = \underset{ \pb_{\theta} }{\arg\max} \left[  \prod_k \theta(y_k | \pb_{\theta})   \prod_l \int_{y_{\rm min}} ^{y_l} \phi(y | \pb_{\theta}) dy \right] \label{eq:ML2}.
\end{align} \\

Here, $N_{x_o}$ and $N_{x_c}$ indicate the number of {\it observed} and {\it censored} $x$ quantities, respectively, whereas $x_{\rm min}$ is the lowest value of $x$ for which the PDF $\phi(x) $ is defined (i.e., $\phi(x) =0$
for $x < x_{\rm min}$). Something similar holds for the quantities $\{ y \}$;
   
\item Computation of the uniform random variates/upper limits  $u_{x_i} = \Phih(x_i)$, $u_{x_j} = \Phih(x_j)$,  $u_{y_k} = \Thetah(y_k), $ and  $u_{y_l} = \Thetah(y_l)$ by means of Eq.~(\ref{eq:Phi})-(\ref{eq:Theta}); \\
\item Computation of the standard Gaussian variates/upper limits  $z_{x_i}$, $z_{x_j}$, $z_{y_k}$ and $z_{y_l}$ by means of Eqs.~(\ref{eq:z1})-(\ref{eq:z2});\\
\item ML estimation of the linear correlation coefficient and then of matrix $\Sigmab$; \\
\item Computation of $\psi(x,y)$ for specific values of $x$ and $y$ by means of Eqs.~(\ref{eq:pdf})-(\ref{eq:cg}).
\end{enumerate}
The {\it copula} related to $z_{x_i}$, $z_{x_j}$, $z_{y_l}$, and $z_{y_k}$ is the same as the one related to $x_i$, $x_j$, $y_k$, and $y_l$. This is due to the {\it invariance property} of {\it copulas} by which
the dependence captured by a {\it copula} is invariant with respect to increasing and continuous transformations of the marginal distributions \citep[see page 13 in][]{TrivediZimmer}.

Such an approach is similar to what is proposed by Takeuchi (2010). 
The difference is that this author estimates $\rho$ and handles the censored data by means of the direct maximisation of the PDF (Eq.~\ref{eq:pdf}), which is
computationally a more expensive method.

\subsection{Application to the luminosity functions}

We apply the procedure above to the case where the BLF has to be estimated for the {\it K-band} and the submm frequencies at $250$, $350$, and $500~{\rm {\mu}m}$.
In this case, 
\begin{align}
\phi(x) &= \phi(L_k) \\
\theta(y) &= \phi(L_{{\rm submm}}).
\end{align}
Before proceeding, it is necessary to fix their analytical form.
Here, it needs to underline that, if $\tau(x)$ is the true PDF  of an experimental quantity $x$ whose measurements are affected by an experimental error $e$ with PDF $\epsilon(e)$, then the PDF that one can estimate is not $\tau(x)$ but
$\phi(x)=\tau(x) * \epsilon(e)$, where ``$*$'' is the convolution operator. In the presence of measurement errors, fitting $\tau(x)$ to the data is not a correct operation. If the measurement errors
are much smaller than the interval of existence of $\phi(x)$,  one could be tempted to assume that the effect of the measurement errors is negligible and therefore that $\phi(x) \approx \tau(x)$. Indeed, this is the case in many practical applications.
However, since $\tau(x+e) \approx \tau(x) + \tau'(x)*e$, with $\tau'(x) = d\tau(x)/dx$, even with small measurement errors, the perturbation term can be important for PDFs with steep slopes. 
The safer procedure for determining $\phi(L_k)$ and $\phi(L_{{\rm submm}})$ is to fit a set of different PDFs to the data and to choose that with the best performance.

We have considered several PDFs to fit the {\it K-band} luminosities. In particular, beside the Schechter's function \citep{Schechter},
which represents the standard analytical form of the LF used in the literature (in practice, it is a {\it gamma} PDF), the Weibull, log-normal and exponential PDFs have been tested. All of them have a support of type 
$L_{\rm min} < L < \infty$, a steep slope for $L \to L_{\rm min}$ and the possibility that $\phi(L) \to \infty$. A point to stress is that, since the quantity $L_{\rm min}$ is unknown, the three-parameter version of such PDF has to be used.
The fit of this kind of PDFs is a difficult problem since the ML approach fails if $\phi(L) \to \infty$ when $L \to L_{\rm min}$. 

\begin{figure}
        \resizebox{\hsize}{!}{\includegraphics{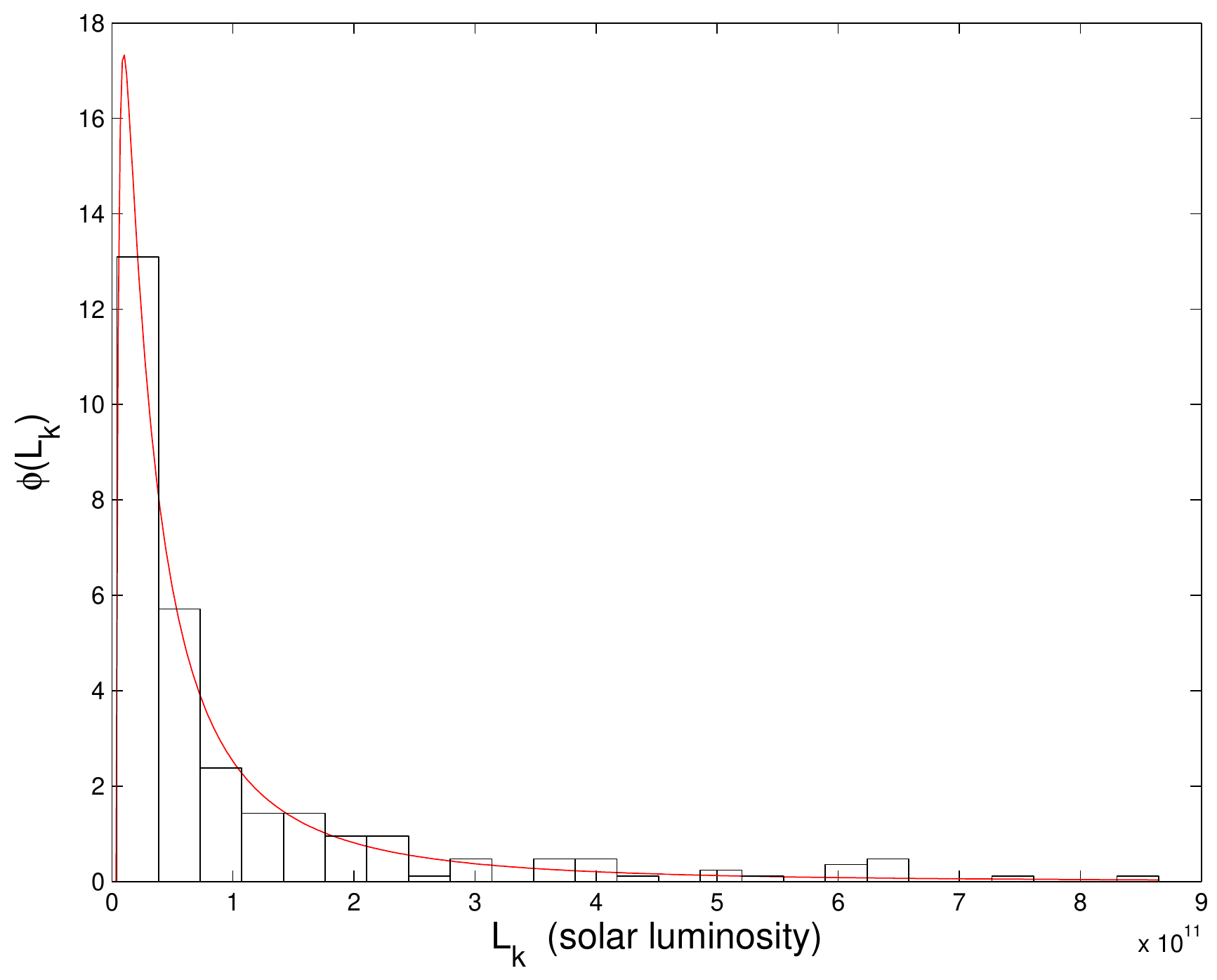}}
        \caption{Histogram of the luminosity $L_k$ in the {\it K-band} for the subsample of late-type galaxies and the corresponding estimated log-normal PDF as obtained with the procedure described in Appendix~\ref{sec:AppendixA}.}
                \label{pdf_k}
\end{figure}

This problem has been solved with  the method described in Appendix~\ref{sec:AppendixA}. 
From this test, the log-normal PDF
\begin{equation}
\phi(L | \mu, \sigma, L_{\rm min}) = \frac{1}{L \sigma \sqrt{2 \pi}} \exp{\left\{ -\frac{[\ln(L-L_{\rm min})-\mu]^2}{2 \sigma^2} \right\}}.
\label{LFopt}
\end{equation}
has proved  to provide the best results for the complete sample of galaxies as well for the subsample of the late-type one (see below).
Here the parameters $\mu$ and $\sigma$ are, respectively, the mean and the standard deviation of the variable $\ln(L-L_{\rm min})$ that, by definition, is normally distributed.
The estimated $\phi(L_k)$ for the late-type galaxies is shown in Fig.~\ref{pdf_k}. 
Thus PDF has been adopted also for the submm bands.

Once the analytical  form of $\phi(L_k) $ and $\phi(L_{{\rm submm}})$ is fixed, the corresponding  parameters are estimated through the maximisation of the quantities~(\ref{eq:ML1})-(\ref{eq:ML2}) with the only constraint that $L_{\rm min} \ge 0$.
The estimated parameters for $\phi(L_k) $ as obtained with this constrained ML method differ less than $2\%$ with respect to those obtained with  the procedure described in Appendix~\ref{sec:AppendixA}.

\section{The sample}
\label{sample}

The HRS is a volume-limited sample (i.e., 15$<D<$25 {\rm Mpc}) that includes late-type galaxies (Sa and later) with 2MASS {\it K-band} magnitude $\leq$12 {\rm mag} and early-type galaxies
(S0a and earlier) with $\leq$ 8.7 {\rm mag}. Additional selection criteria are high Galactic latitude ($b >+55^\circ$)
and low Galactic extinction (AB$<$0.2 {\rm mag}, \citep{Schlegel+98}).
The sample includes 322 galaxies (260 late and 62 early types). The selection criteria are fully described in Boselli et al. (2010a). 
The volume covered by the HRS is calculated considering that, according to the selection criteria adopted to define the sample \citep{Bose+10a}, we selected
galaxies in the volume between 15 and 25 Mpc over an area of 3649 sq.deg., which leads to a volume of 4539 Mpc$^3$. 
Morphological types and distances are taken from Cortese et al. (2012a). 
Briefly, we fixed the distances
for galaxies belonging to the Virgo cluster (23 Mpc for the Virgo B cloud and
17 Mpc for all the other clouds, Gavazzi et al. 1999), while for the rest of the sample ,distances were
estimated from their recessional velocities assuming a Hubble constant H$_{0}$ = 70 {\rm km s}$^{-1}$ {\rm Mpc}$^{-1}$.
Additional information about this sample may be found in Boselli et al. (2010a) and Cortese et al. (2012a). 

The HRS was targeted as part of the {\textit{Herschel}}/SPIRE \citep{Griffin+10} guaranteed time
\citep{Bose+10a}. In this paper, we use the submm fluxes observed by the SPIRE photometer at 250, 350, and 500 {\rm $\mu$m}  as
published in Ciesla et al. (2012). 

The sample has neither a homogeneous selection at the {\it K-band} nor a homogenous cut at submm luminosity. Observations of early-types reached a 4 times lower sensitivity limit; i.e., we treat the early-type sources as though they were observed to
approximately the same sensitivity limit as the late-type sources, and it does not include star-forming low-mass galaxies that would have
large submm but low {\it K-band} emissions.
Therefore, by construction, the HRS is {\it not} complete at submm wavelengths. 
Part of the SPIRE observations did not turn out to be detections and mostly among early-type galaxies, the estimated submm fluxes are upper limits (see also Smith et al 2012a).
These upper limits depend on the absolute luminosity, at least for objects with low luminosity.

The sample has a very limited luminosity coverage, the maximum observed luminosity at 250$\mu$m is $10^9$L$_\odot$, and it contains the Virgo cluster, which might introduce two biases: (1) Clusters are dominated by early types with respect to
field galaxies (the morphology segregation effect, \citep{Dressler80}). The HRS sample therefore contains a significant fraction of early-type galaxies much more than one would normally find in a ``blindly generated" sample (such 
the H-ATLAS in Vaccari et al. (2010), where the portion of cluster galaxies is only a few percent).
(2) The late-type galaxies in the cluster are different from field galaxies for multiple reasons. For instance they have reduced star formation and therefore a reduced far-infrared emission
because they are poorer in gas \citep{BoseGava06}. Recently, Cortese et al (2010, 2012a) have shown that the HRS late-type galaxies in clusters have truncated dust discs and lower dust masses.
This might introduce a non-homogenous K-mag distribution for late type galaxies because of the presence of two types of late-types: cluster and field galaxies.


\section{Results}\label{results}
\subsection{The {\it K-band} LF for the HRS sample}

The {\it K-band} LF of the 2MASS sample (flux-limited over a large sky volume), from which the HRS has been extracted, has already been determined by Kochanek et al. (2001) and Cole et al. (2001). 
As a first check we plot  the values of the {\it K-band} LF estimated by Kochanek et al. (2001) in Fig.~\ref{LF-K} and overlap those found for the HRS, using the simple test of counting galaxies within the luminosity bins.
As already shown in Boselli et al (2010), the {\it K-band} LF computed on the HRS sample agrees within the errorbars with that of the parent sample within the luminosity range sampled by the HRS, although the HRS only spans a very limited range of $L$ (see section~\ref{sample}).


We use the lognormal form of the LF shown in equation (\ref{LFopt}) to fit the {\it K-band}, 250, 350, and 500$\mu$m LF with the same procedure as for all frequencies, as outlined in Appendix A.
It is worth stressing here that the goal of this work is not to try to find the LF best-fit parameters and, as shown in Appendix~\ref{sec:AppendixA}, it would be equally good to fit the standard Schechter functions. Function~(\ref{LFopt}) {\it only} provides a better description of the LF of the {\it present} sample.



\subsection{The bivariate {\it K-band} /submm LF}

\begin{figure}
        \resizebox{\hsize}{!}{\includegraphics{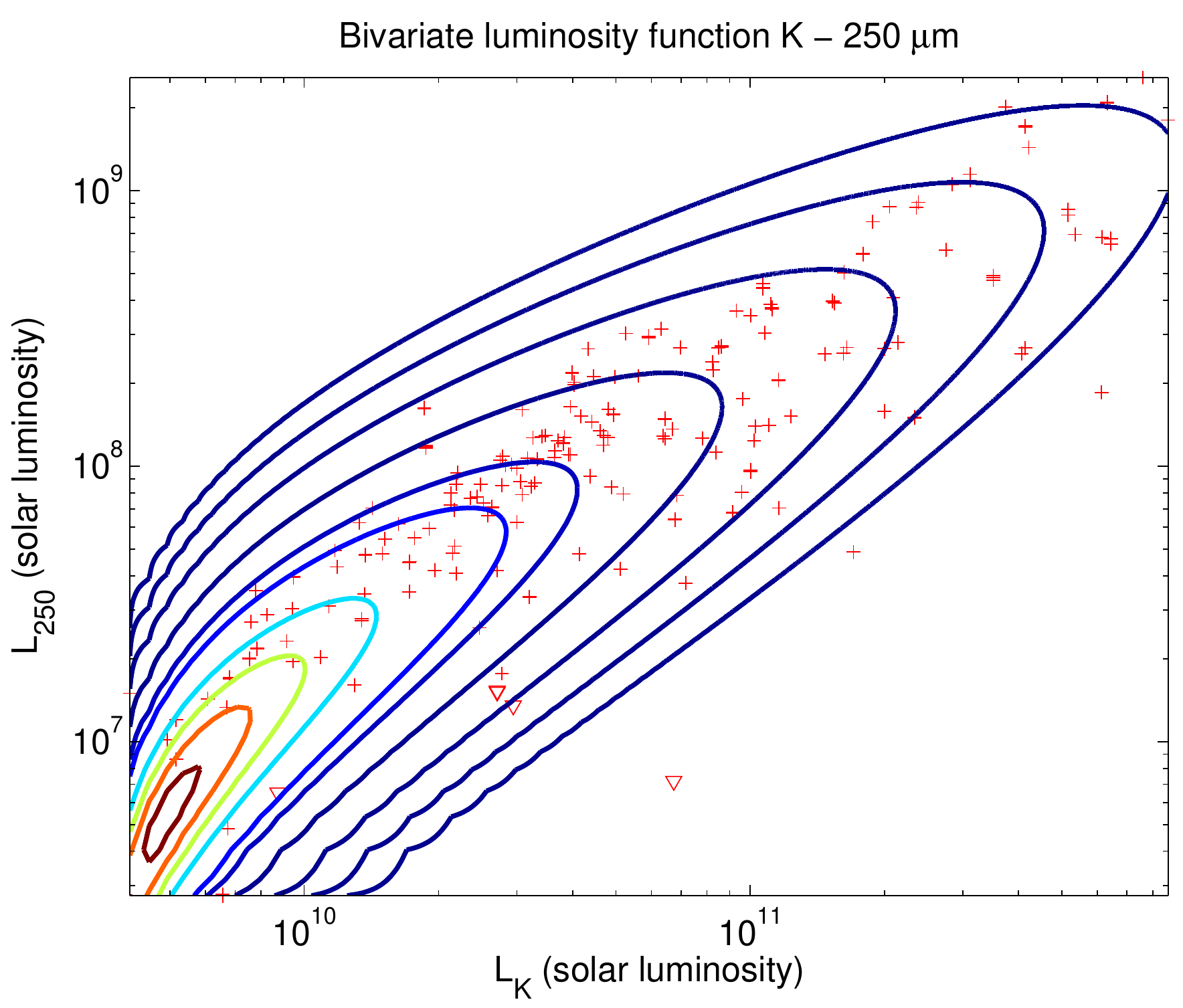} 
        }
        \caption{Estimated bivariate PDF shown on logarithmic scale for the ${\rm K}$ band with the $250~{\rm{\mu}m}$-band, for only late-type galaxies. Crosses correspond to the detected fluxes, while downward-pointing triangles to upper limits.
        Contour lines correspond to the levels 0.00001, 0.0001, 0.001, 0.01, 0.05, 0.1, 0.3, 0.5, 0.7, 0.9. These values correspond to the fraction of the peak value of the BLF  that is set to one. The figure clearly shows the correlation among the data.}
        \label{fig:bipdfslog-250}
        \end{figure}

\begin{figure}
        \resizebox{\hsize}{!} 
        {\includegraphics{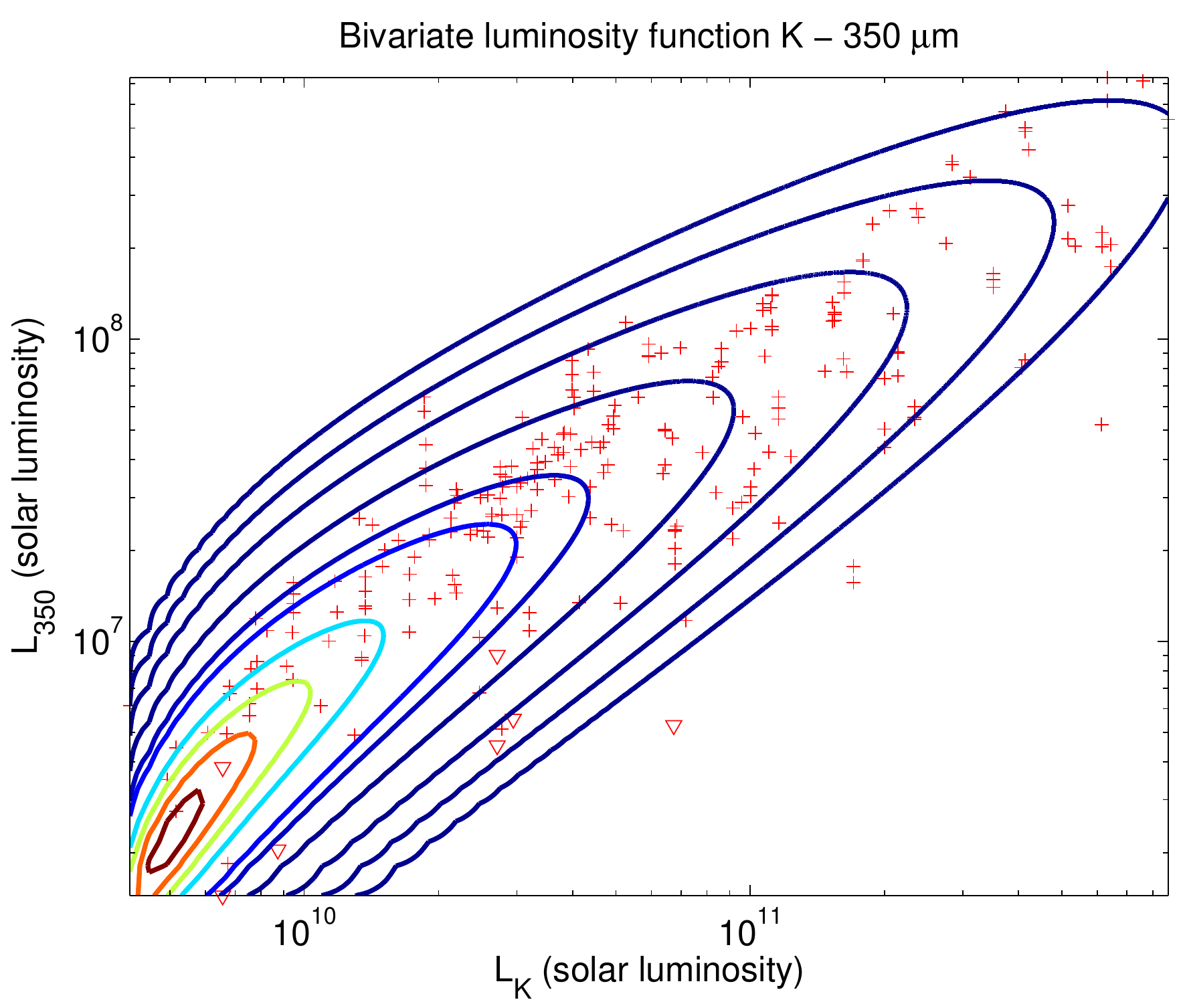}
        }
        \caption{As in Figure~\ref{fig:bipdfslog-250} for the $350~{\rm{\mu}m}$-band.}
        \label{fig:bipdfslog-350}

\end{figure}

\begin{figure}
        \resizebox{\hsize}{!}{\includegraphics{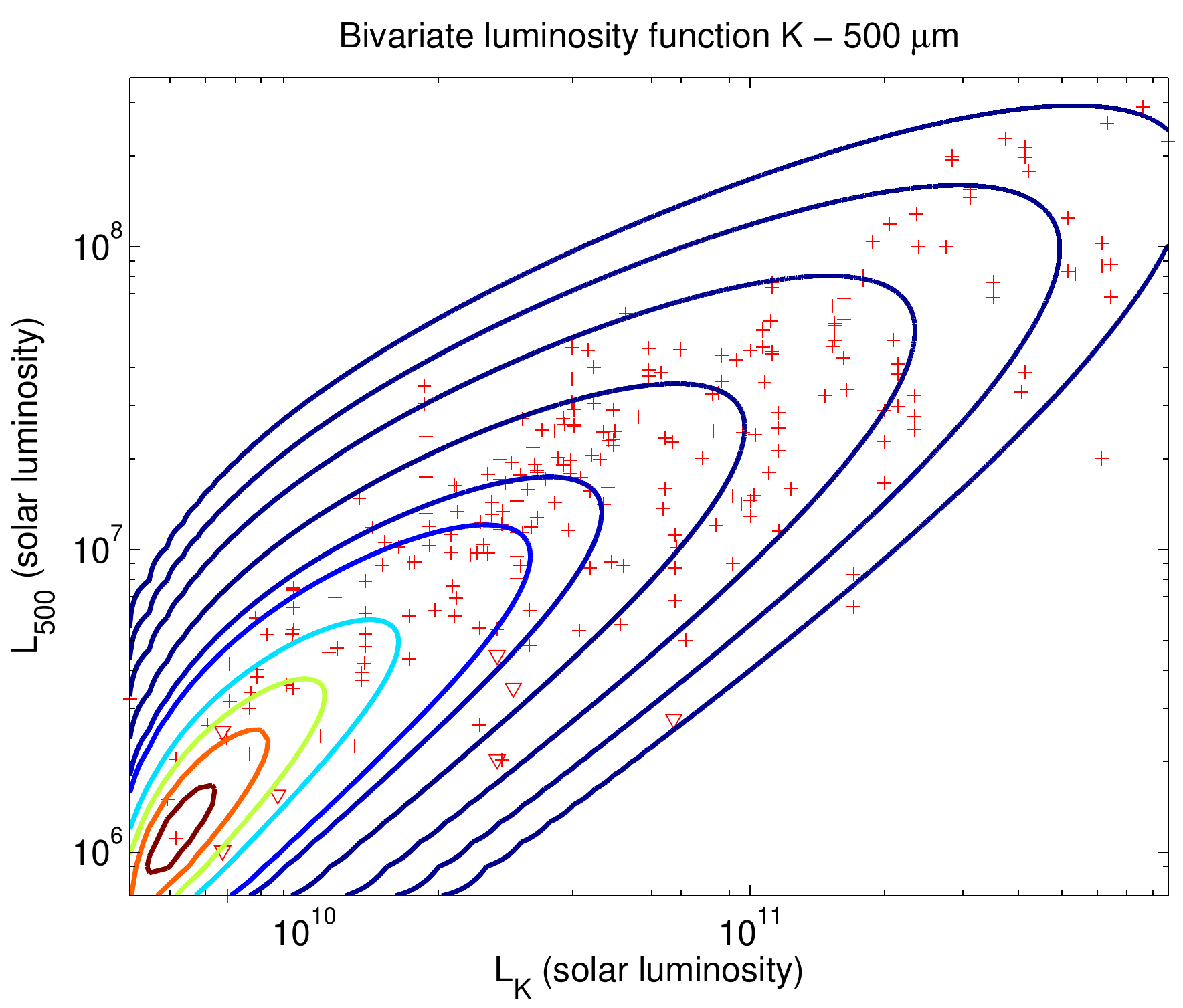}}
        \caption{As in Figure~\ref{fig:bipdfslog-250} for the $500~{\rm{\mu}m}$-band}
        \label{fig:bipdfslog-500}
\end{figure}

We first applied the procedure outlines in Sect.\ref{tools} to the late-type objects that constitute more than 75\% of the total number of sources in the sample (260 out of 323) \citep{cor+12a}.
Figures~\ref{fig:bipdfslog-250}-\ref{fig:bipdfslog-500} show the estimated bivariate PDF of the luminosity in the  {\it K-band} with the luminosity corresponding to the various submm frequencies on logarithmic scale.

A strong relationship between the {\it K-band} and the various submm frequencies is evident from these figures. 
This impression is confirmed by the linear correlation coefficient $\rhoh$ that is $0.93$, $0.93$, and $0.91$ 
between the Gaussian variates corresponding to the  {\it K-band} and those corresponding to the $250$, $350$ and $500~{\mu}m$ 
frequencies (step 4 of Sect.~\ref{tools}). Given how few early-type galaxies there are (62 objects), the same procedure is unable to provide stable results. 

\begin{figure}
        \resizebox{\hsize}{!}{\includegraphics{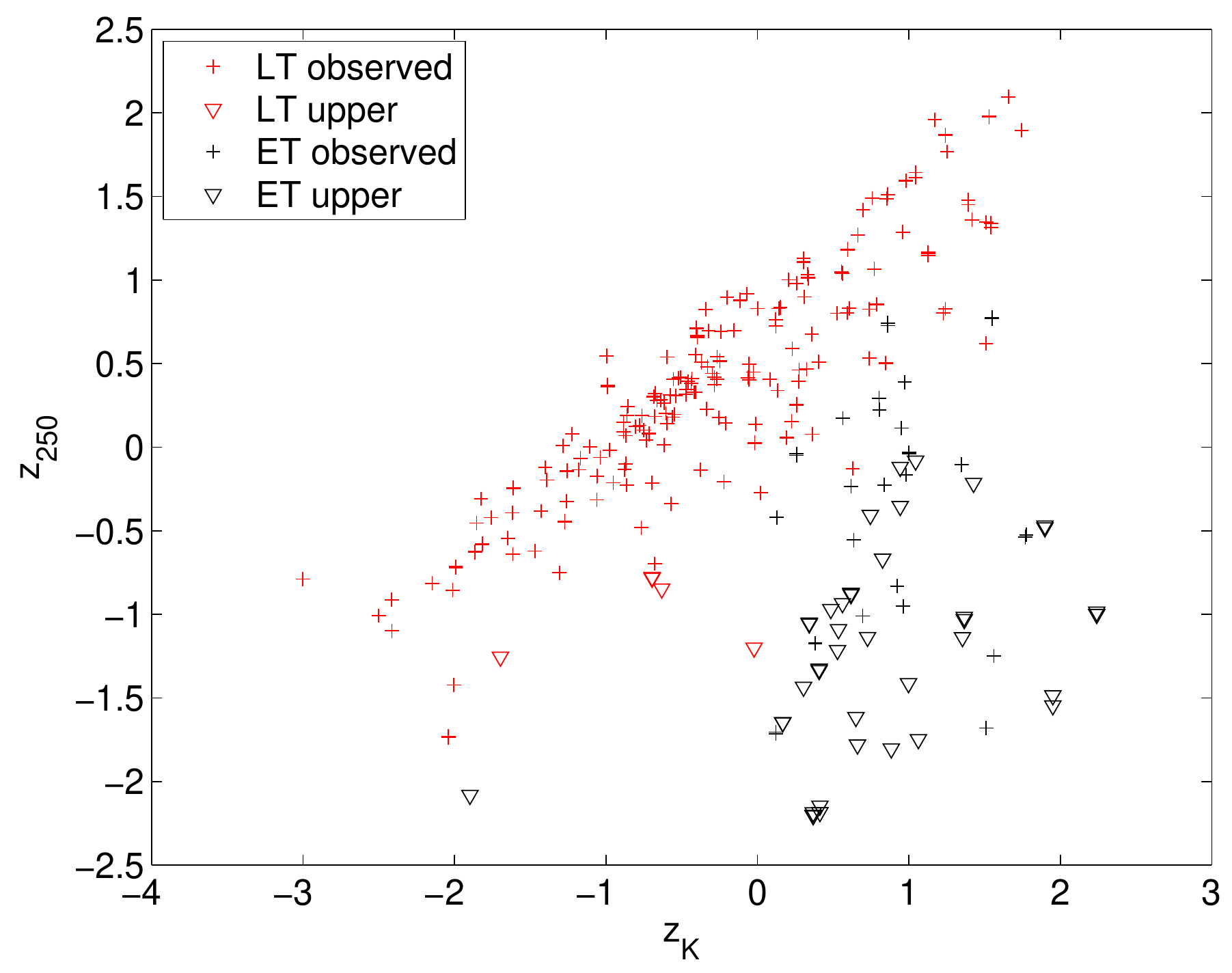}}
        \caption{Bivariate Gaussian distribution of the  {\it Gaussianised}  HRS data: {\it K-band} versus 250$\mu$m data. Red symbols correspond to late-type galaxies, black ones to early-type. Upper limits are identified as downward triangles.
	Here, $z = G^{-1}(u)$ with $G^{-1}(u)$ the inverse function of the standard Gaussian CDF, and $u$ is the uniform random variable from the CDF of the LF at the given frequency (see text).}
        \label{BG250}
        
\end{figure}

\begin{figure}
        \resizebox{\hsize}{!}{\includegraphics{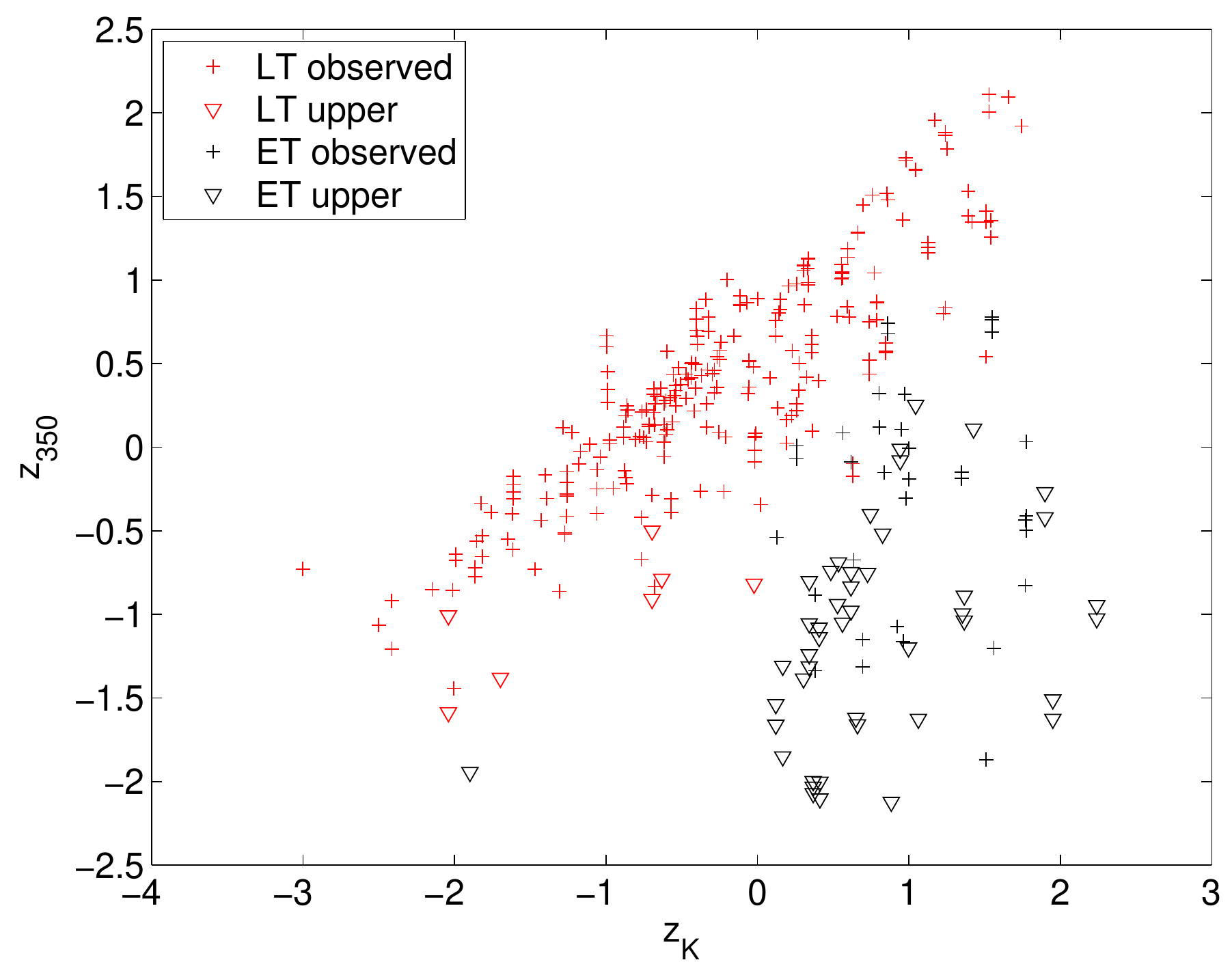}}
        \caption{As in Figure~\ref{BG250} but for the 350$\mu$m data.}
        \label{BG350}
\end{figure}

\begin{figure*}
\centering
   \includegraphics[width=17cm]{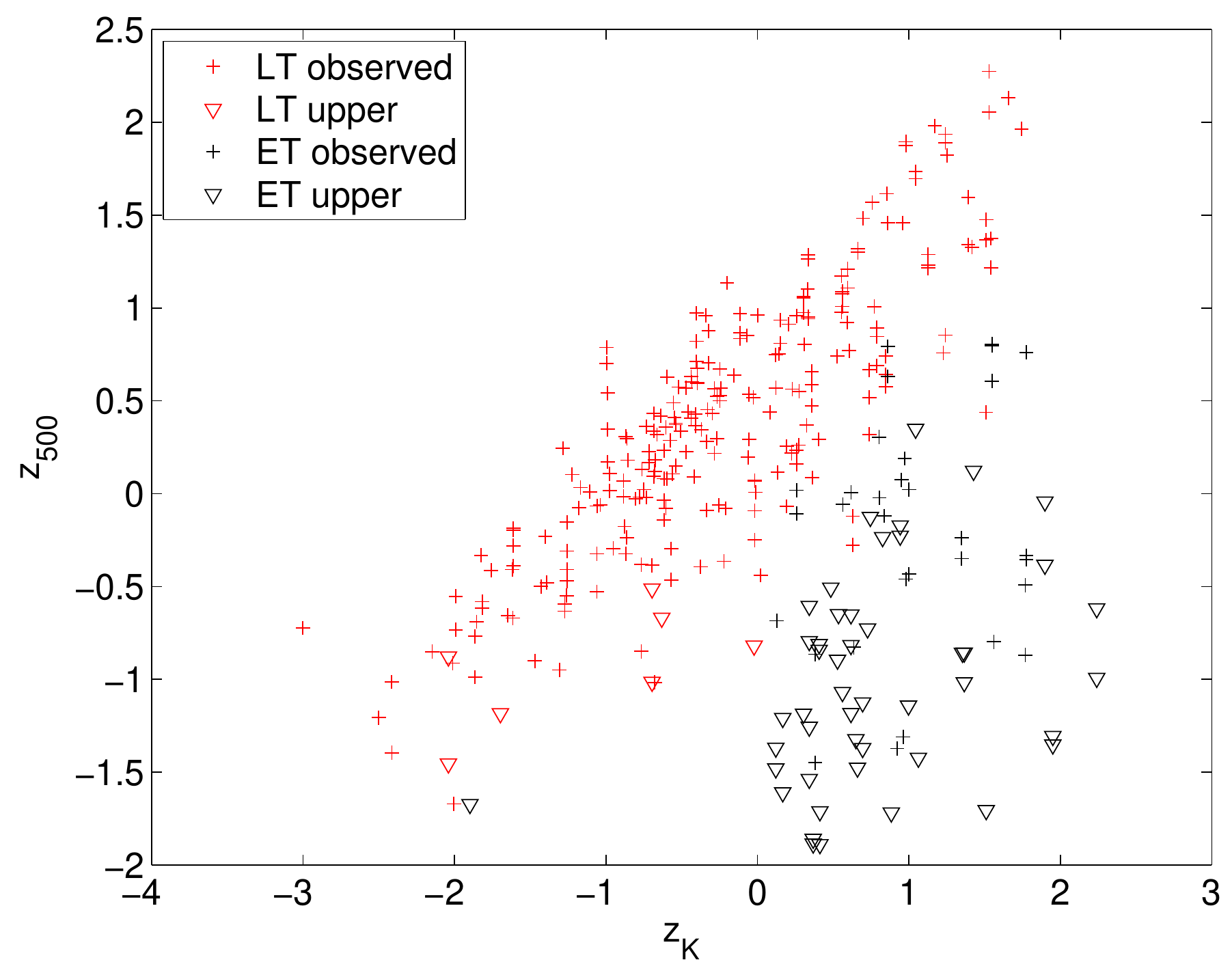}
     \caption{Same as Figure~\ref{BG250} but for the 500$\mu$m data.}
     \label{BG500}
\end{figure*}

When the above procedure is applied to the whole  HRS sample, the linear correlation coefficient $\rhoh$ becomes $0.66$, $0.67$ and $0.69$, respectively. An explanation of this fact can be inferred from Figs.~\ref{BG250}-\ref{BG500} that show the bivariate Gaussian distributions for the three sets [{\it K-band}, 250$\mu$m]
 [{\it K-band}, 350$\mu$m] [{\it K-band}, 500$\mu$m] after the operation of {\it Gaussianisation} of the luminosities. An inspection of these figures clearly highlights that distribution of the data does not conform to a bivariate Gaussian. The point
is that this procedure is based on the assumption that, for a given frequency, the luminosity of all the object shares the same PDF. However, from Figs.~\ref{BG250}-\ref{BG500} it is evident that the early-type and the late-type galaxies
have different PDFs. To quantify this impression, we tested the hypothesis that the early type galaxies share the same PDF of the late type galaxies by means of a Gehan's generalised Wilcoxon test. Such a test permits comparing two sets of data in the presence of censored observations \citep{Lee}. For all frequencies, this hypothesis is rejected at the 0.001 confidence level.
The conclusion is that a tight relationship between {\it K-band} and submm luminosities is not valid over the whole HRS sample.



\begin{figure}
        \resizebox{\hsize}{!}{\includegraphics{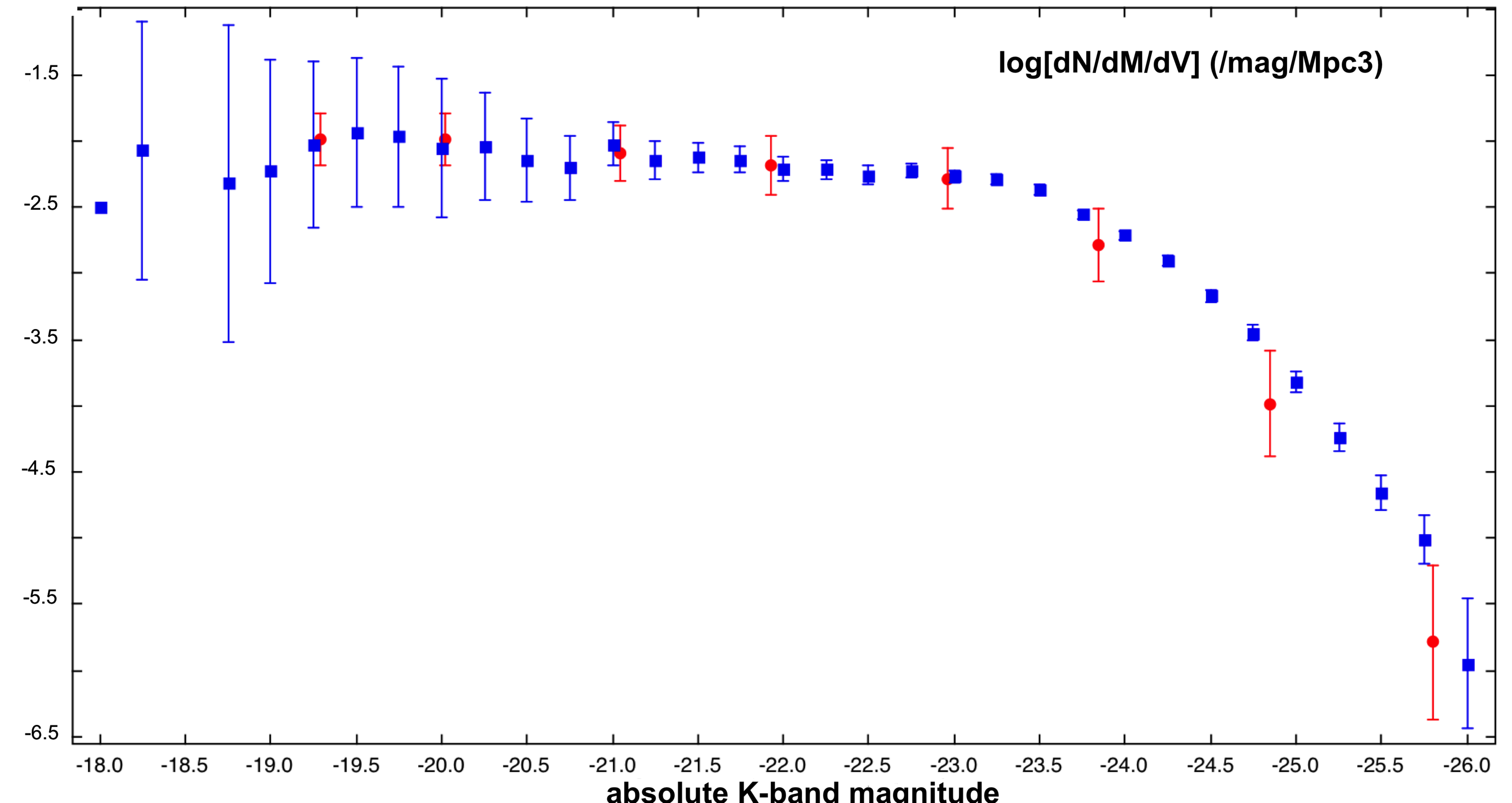}}
        \caption{The {\it K-band} LF computed for the HRS sample (red circles). Superimposed are the values of the {\it K-band} LF computed by Kochanek et al. (2001) over the whole 2MASS sample.}
                \label{LF-K}
\end{figure}

The statistical analysis used here is based on the fact that the present sample -- as discussed in Sect.~\ref{sample} above -- can be considered to be representative of the parent {\it K-band} (homogenous) sample (see Fig.~\ref{LF-K}). The incompleteness of the sample in the submm (not blindly selected in submm) and the above assumption allowed us to apply the algorithms of the BLF. The additional fact that the sample is volume-limited makes us state that the results presented here are not affected by the incompleteness, and we may furthermore say that the assumption that all the galaxies seen in the {K-band} are also seen in the submm bands is a valid one.

\section{Discussion}\label{discussion}

A statistical analysis of the whole HRS sample, which takes the presence of upper limits among the flux estimations into account, has shown that no statistically meaningful results can be obtained regarding the linked physical properties as traced by photometric observations in the {\it K-band} and those traced by the submm ones. However, we find that the analysis being restricted to a subsample of late-type galaxies highlights the well constrained behaviour of the BLF.

For late-type galaxies, the estimation of the BLF shows that the {\it K-band} and the submm bands are tightly correlated, and the BLF provides all the information on the distribution of the two luminosities and their relation.
One could argue that since all luminosities are calculated using distance squared, this relation is artificial, especially in logarithm space.  However, the
HRS sample is limited in distance (in which galaxies were selected if they fell between an inner and outer distance limit), and the distance-related bias (the so-called {\it Malmquist bias}) does not play any role.
The relation between {\it K-band} and submm therefore also  corresponds to the one between fluxes and apparent magnitude. This result suggests that for late-type galaxies the brighter the galaxy in the {\it K-band}, the brighter it is in the submm. The locus of the figures at low {\it K-mag} is not populated due to a selection effect
whereby we do not observed the low surface brightness galaxies;, i.e. galaxies exist but are not represented in these figures \citep{Bose+10a}.

The submm luminosity, at least for nearby late-type galaxies, is mainly due to the cold component of the dust that is closely
related to the galaxy dust mass, and the stellar mass can be inferred from the {\it K-band} absolute luminosity, which is indeed a tracer of the mass of the old stellar population. The tight relationship between {\it K-band} and submm emission can be interpreted as relationships among the stellar mass, the cold dust mass, and the far-infrared luminosity in late-type galaxies.

Our result agrees with what was found by Bourne et al. (2012) for a much larger sample of optical galaxies observed by SPIRE within the H-ATLAS survey. For nearby objects, these authors show a tight relation between the SPIRE fluxes and the {\it red} luminosity.
A physical link between the stellar mass with the dust luminosity and the cold dust mass has been suggested by Cortese et al. (2012b),
Bourne et al. (2012), Agius (2013) and Clemens (2013). 
These authors find a general trend toward a decreasing ratio between the dust and stellar mass, $M_{\rm dust}/M_\star$, with the stellar mass in a variety of nearby objects. For massive galaxies, the dust content with respect to the mass in the old stars is reduced and lower mass galaxies have a higher dust fraction. Clemens et al. (2013) interpret this finding further and claim that lower mass galaxies have more dust because they have proportionally more interstellar medium and that the star formation rate is mainly determined from the dust mass of a galaxy and not from the stellar mass within the stellar masses sampled in the nearby Universe. 
An additional mechanism linking the stellar mass to the dust seen at submm wavelengths is the heating by the interstellar radiation field from the
total stellar population (including evolved stars), which is primarily seen at NIR wavelengths \citep[e.g.][]{Bendo+10,Boquien+11,Bendo+12,Groves+12,Smith+12b}. 

These findings may explain the observed shape of the BLF. The slight scatter of the BLF, or its shape bending towards large {\it K-band} luminosities and lower submm ones is naturally reproduced by the {\it copula} BLF and can be quantified.
There is a tight relation at lower luminosities, while the spread is greater at larger luminosities where we observe a drop in the value of the correlation coefficient.
This could be explained as another physical mechanism intervening and contributing to the submm luminosities in sources with higher luminosities, which is a higher dust content in galaxies of lower stellar mass. For instance a warmer dust component, whose spectral energy distribution would peak at shorter wavelength would have a lower flux at submm. Boselli et al. (2010b and 2012) indeed find that the shape of the spectral energy distributions between 250 and 500{\rm $\mu$m} changes significantly from metal-poor and/or star-forming galaxies and quiescent spirals.

The situation is much more controversial for early-type galaxies.  On the one hand, the trend seen for late-type galaxies is followed by early-type galaxies, too, although with a factor of ten lower normalisation owing to the overall reduced dust content in these galaxies.
On the other hand, they also show independence of the {\it K-band} with respect the SPIRE bands luminosities which may be linked to cold dust not sharing the physical location of the old star population.
The reason the {\it K-band} and submm flux densities may not be well-correlated is probably that the dust mass within galaxies is
related to the stellar disc mass rather than the global stellar mass.
Since early-type galaxies have relatively small or negligible discs, the stellar mass appears uncorrelated with dust emission. 

It is common to find that dust has a different distribution from the stellar population in both early- and late-type galaxies.
Boquien et al. (2011), Bendo et al. (2012), Groves et al. (2012), and Smith et al. (2012b) all show that it is even possible for dust to be heated by evolved stars
in a large bulge even if the dust and stars have different distributions (as is the case in M31 and M81) and that the temperature of the cold dust is tightly driven by the evolved stellar populations. 

Mathews et al. (2013) suggest that this lack of dependency is due to the relocation of variable masses of cold dust away from the central reservoir, eventually triggered by the central AGN and then destroyed by sputtering \citep[see also][]{Smith+12a,diSerego+13}. 
Indeed, among the HRS early-type galaxies, M86 shows dust that is displaced with respect to the nucleus, and probably accreted from an external object during a merging process \citep{Gomez+10}, M87 whose submm emission is mainly due to synchrotron \citep{Baes+10}, and M84 whose 500$\mu$m and partially likely 350$\mu$m
emissions are dominated by synchrotron radiation \citep{Bose+10b}.

The results presented here are pertinent to a well defined local sample of objects, are valid in the nearby Universe and might be not applicable to galaxies at larger redshifts. For instance, the low temperature of the dust may be related to the quiescent status of the star formation activity in local galaxies, while at higher redshift the dust is heated at higher temperature from the more copious massive stars, and the relationship between the cold dust mass and the galaxy stellar mass might not hold any longer.
Indeed, Santini et al (2013) find that the relationship between dust mass and stellar mass is not longer valid for galaxies at higher redshift.

\section{Conclusions}
The relationship between the {\it K-band} and the far infrared (submm) emissions of nearby galaxies, investigated through the BLF [{\it K-band}, submm] of the {\textit{Herschel}} Reference Survey (HRS), shows that while a poor correlation is found using a whole HRS sample.
The late-type galaxies subsample presents a remarkable correlation, which reflects on the physical relationship between the two frequency ranges.
These results suggest that the LF of galaxies computed in the {\it K-band} and in the submm are dependent and the dependence lies in the relationship between the galaxy
stellar mass and the cold dust mass.

Finally, the agreement for the estimate of the LF of the HRS sample with what was estimated for the parent sample (2MASS survey)
further supports the conclusion that inhomogeneities associated with large scale structure in the local Universe does not affect the  statistical properties
of the galaxy distribution much.
This conclusion, together with our new statistical analysis, has allowed us to better characterise the HRS sample, where, despite the in-homogenous selection and a partial bias towards low submm luminosities, our sample can reasonably be considered as representative of a nearby population of galaxies.

\begin{acknowledgements} 
Herschel is an ESA space observatory with science instruments provided by European-led Principal Investigator consortia and with important participation from NASA.
PA is grateful to the IAPS Institute for hospitality during the submmst part of this work, and to ESO for support through the DGDF fund of 2012.   
IDL is a postdoctoral researcher of the FWO-Vlaanderen (Belgium).
\end{acknowledgements}





\clearpage
\appendix

\section{Testing for goodness fit of three-parameter distributions} \label{sec:AppendixA}

\begin{figure}
        \resizebox{\hsize}{!}{\includegraphics{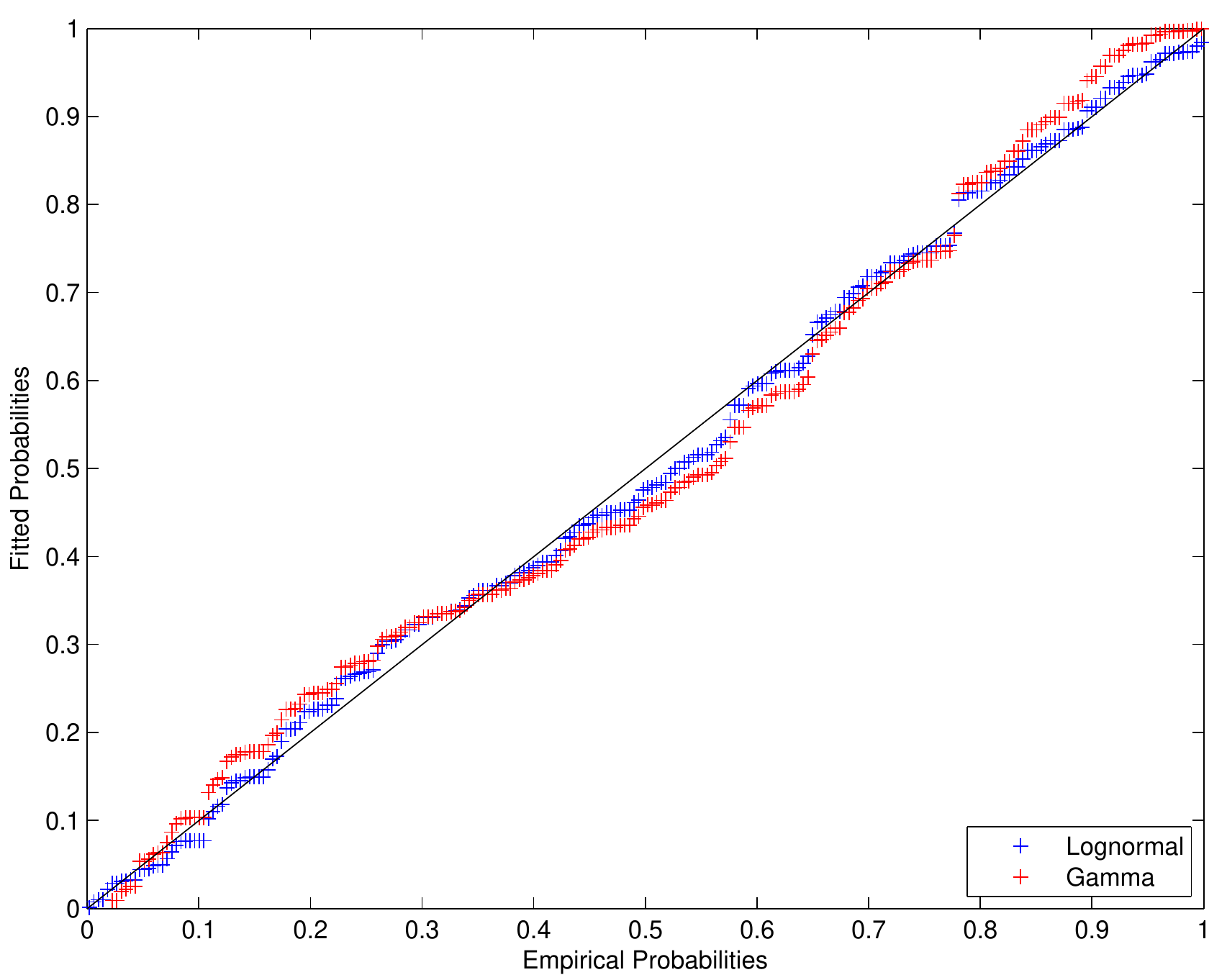}}
        \caption{Goodness test for the three parameter log-normal and Schechter probability density function for the late type galaxies in the {\it K-band}.}
        \label{fig:test}
\end{figure}

When fitting a probability density function (PDF) to a set of $N$ data $\{ x_i \}_{i=1}^{N}$, two parameters often have to be estimated that typically are linked to its position and shape 
(e.g. in the case of the Gaussian PDF these parameters are the mean and the 
standard deviation). There are situations, however, where the PDF, $\phi(x)$ is defined in a domain $ x_{{\rm min}} - \infty$, with $x_{{\rm min}}$ an unknown quantity; i.e., the parameters to estimate are three. In this case,
the method of ML often fails to provide useful parameter estimates. For example,
this happens with PDFs, such as the Schechter one, for which $\phi( x_{{\rm min}}) = \infty$. A possible way out is an approach based on the use of the empirical cumulative distribution function $\widehat{\Phi}(x_i)$ (ECDF) given by
\begin{equation}
\widehat{\Phi}(x_i) = \frac{i-0.5}{N}
\end{equation}
with $\{ x_i \}$ sorted in increasing order. The idea is that, if an ECDF is plotted
versus the values of the CDF $\Phi(x_i)$ corresponding to the true PDF $\phi(x_i)$, then the resulting points distribute along a straight line with unit slope. An estimate of the parameters can therefore be obtained from the PDF  whose CDF 
provides the ECDF-CDF point distribution closest, in the least-squares sense, to such line. In practice, the three parameters are iteratively changed, the corresponding $\Phi(x_i)$ evaluated and finally the sum of the squared distances of the  
ECDF-CDF point distribution from the straight line computed. In this operation, it is useful to weight the data to give more importance to the tails of the PDF that are more able to distinguish the various types.
The weights $\{ w_i \}$ can be defined in terms of $\widehat{\Phi}(x_i)$
\begin{equation}
w_i = \frac{1}{ \sqrt{\widehat{\Phi}(x_i) \left[ 1-\widehat{\Phi}(x_i) \right]} }.
\end{equation}
Figure \ref{fig:test} shows the results for the data in the {\it K-band} for the sample of late-type galaxies when this method is applied with two PDFs given by a log-normal and a Schechter distribution. The superiority of
the fit provided by the former is evident. In particular, the root-mean square of the distances for the log-normal PDF is $1.1 \times 10^{-3}$ vs. $2.1 \times 10^{-3}$ for the Schechter one.

\begin{thebibliography}{}
\bibitem[\protect\citeauthoryear{Agius et al. 2013}{Agius et al.}{2013}]{Agius} Agius N.K. et al, 2013, MNRAS 431, 1929
\bibitem[Baes et al. 2010]{Baes+10} Baes M., Clemens M., Xilouris E. M., Fritz J., Cotton W. D., Davies J. I., Bendo G. J., Bianchi S., Cortese L., de Looze I.et al 2010, A\&A 518, L5
\bibitem[Ball et al. 2006]{Bal+06} Ball N.M., Loveday J., Brunner R.J., Baldry I.K., Brinkmann J., 2006 MNRAS 373, 845
\bibitem[Bendo et al. 2010]{Bendo+10} Bendo G.J. et al., 2010, A\&A 518, L65
\bibitem[Bendo et al. 2012]{Bendo+12} Bendo G.J. et al., 2012, MNRAS 419, 1833
\bibitem[Boquien et al. 2011]{Boquien+11} Boquien, M.; Calzetti, D.; Combes, F.; Henkel, C.; Israel, F.; Kramer, C.; Rela\~no, M.; Verley, S.; van der Werf, P.; Xilouris, E. M.; HERM33ES Team, 2011, AJ 142, 111
\bibitem[Boselli et al. 2010a]{Bose+10a} Boselli A. et al., 2010a, PASP 122, 261 
\bibitem[Boselli et al. 2010b]{Bose+10b} Boselli A. et al., 2010b, A\&A 518, L61
\bibitem[Boselli 2011]{Bose11} Boselli A. {\it A Panchromatic View of Galaxies}, Practical Approach Book, Wiley-VCH, Berlin 2011
\bibitem[Boselli \& Gavazzi 2006]{BoseGava06} Boselli A., Gavazzi G., 2006 PASP 118, 517
\bibitem[Bourne et al. (2012)]{Bourne+12} Bourne N., et al., 2012, MNRAS 421, 3027
\bibitem[Chapman et al 2003]{Chap+03} Chapman S.C., Helou G., Lewis G.F., Dale D.A. 2003, ApJ 588, 186
\bibitem[Choloniewski 1985]{Cholo85} Choloniewski J., 1985, MNRAS 214, 197
\bibitem[Ciesla et al. (2012)]{cies+12} Ciesla L. et al., 2012, A\&A 543, 161
\bibitem[Clemens et al.]{clemens} Clemens M.S. et al., 2013, MNRAS 433, 695 
\bibitem[Cole et al. 2001]{Cole+01} Cole S., et al., 2001, MNRAS, 393, 681
\bibitem[Cortese et al. (2012)]{cor+12} Cortese L., et al., 2010, A\&A 518, L49
\bibitem[Cortese et al. (2012b)]{cor+12b} Cortese L., et al., 2012b, A\&A  544, 101 
\bibitem[Cortese et al. 2012a]{cor+12a} Cortese L., et al., 2012a, A\&A 540, 52 
\bibitem[Cross \& Driver 2002]{CrossDriver02} Cross N., Driver S.P., 2002 MNRAS 329, 579
\bibitem[Davies et al. 2010]{Davies+10} Davies J. I., Baes M., Bendo G.J., Bianchi S., Bomans D.J., Boselli A., Clemens M., Corbelli E., Cortese L., 
Dariush A. et al., 2010 A\&A 518, L48 
\bibitem[di Serego Alighieri et al. 2013]{diSerego+13} di Serego Alighieri S., et al., 2013, A\&A 552, 8
\bibitem[Dressler 1980]{Dressler80} Dressler A., 1980, ApJ 236, 351
\bibitem[Driver et al. 2006]{Driver+06} Driver S.P. et al., 2006 MNRAS 368, 414
\bibitem[Dunne et al. 2011]{Dunne+11} Dunne L., Gomez H. L., da Cunha E., Charlot S., Dye S., Eales S., Maddox S.J., Rowlands K., Smith
D.J.B., Auld R. et al., 2011, MNRAS 417, 1510
\bibitem[Dye et al. 2010]{Dye+10} Dye S., et al., 2010 A\&A 510, L10
\bibitem[\protect\citeauthoryear{Efstathiou, Ellis, Peterson (1988)}{Efstathiou et al. 1988}]{Efstathiou88}  Efstathiou G., Ellis R.S., Peterson B.A.
1988 MNRAS 232, 431	
\bibitem[Gavazzi et al., 1996]{gavazzi96} Gavazzi G. et al 1996, A\&A 312, 397
\bibitem[Gavazzi et al. 1999]{gavazzi} Gavazzi G., Boselli A., Scodeggio M., Pierini D., Belsole, E. 1999, MNRAS 304, 595
\bibitem[Gomez et al. 2010]{Gomez+10} Gomez H. L., Baes M., Cortese L., Smith M. W. L.,  Boselli A., Ciesla L., Bendo G. J. et al 2010, A\&A, 518, L45
\bibitem[Griffin et al. 2010]{Griffin+10} Griffin, M.J., Abergel, A., Abreu, A. et al. 2010, A\&A, 518, L3
\bibitem[Groves et al. 2012]{Groves+12} Groves B. et al., 2012, MNRAS 426, 892
\bibitem[Johnston 2011]{Johnston11} Johnston R., 2011, A\&ARv 19, 41
\bibitem[Kochanek et al. 2001]{Kochanek+01} Kochanek C.S., Pahre M.A., Falco E.E., Huchra J.P., Mader M., Jarrett T.H., Chester T., and Cutri R., Schneider S.E.
2001, ApJ 560, 566
\bibitem[La Franca et al. 1995]{LaFranca+95} La Franca F., Franceschini A., Cristiani S.,Vio R., 1995 A\&A 299, 19
\bibitem[Lee 1992]{Lee} Lee E.T. 1992, Statistical Methods for Survival Data Analysis, John Wiley \& Sons: New York
\bibitem[Mathews et al. 2013]{Mathews+13} Mathews W.G., Temi P., Brighenti F., Amblard A. 2013, ApJ 768, 28
\bibitem[Pilbratt et al. 2010]{herschel} Pilbratt, G.L., Riedinger, J.R., Passvogel, T. et al. 2010, A\&A 518, L1
\bibitem[Poglitsch et al. 2010]{PACS} Poglitsch, A., Waelkens, C., Geis, N. et al. 2010, A\&A 518, L2
\bibitem[Santini et al. 2013]{Santini+13} Santini P., 2013, arXiv1311.3670S
\bibitem[Saunders et al 2010]{Saunders+10} Saunders, W., Rowan-Robinson, M., Lawrence, A., Efstathiou, G., Kaiser, 
N., Ellis, R. S., Frenk, C. S., 1990, MNRAS 242, 318
\bibitem[Schafer 2007]{Schaf07} Schafer C.M., 2007, ApJ 661, 703
\bibitem[Schechter 1976]{Schechter} Schechter, P., 1976, ApJ  203, 297, 1976
\bibitem[Schlegel et al. 1998]{Schlegel+98} Schlegel, D.J., Finkbeiner, D.P., Davis M., 1998 ApJ Supplement 500, 525
\bibitem[Schmidt 2007]{Schmidt07}  Schmidt T., in {\it Copulas: From Theory to Application in Finance}, J\"orn Rank (Editor), published by Risk Books, 2007
\bibitem[Smith 2012]{SmithR12} Smith R.E., 2012, MNRAS 
\bibitem[Smith et al. 2012a]{Smith+12a} Smith M.W.L. et al., 2012a, ApJ 748, 123
\bibitem[Smith et al. 2012b]{Smith+12b} Smith M.W.L. et al., 2012b, ApJ 756, 40
\bibitem[Takeuchi 2010]{Tak+10} Takeuchi T. T., 2010, MNRAS 406, 1830
\bibitem[Trivedi and Zimmer 2005]{TrivediZimmer} Trivedi P.K. \& Zimmer D.M., 2005, {\it Copula modelling: an introduction for practitioners}, now Publishers Inc., Hanover: USA
\bibitem[Vaccari et al. 2010]{Vac+10} Vaccari M. et al., 2010, A\&A 510, 20
\end{thebibliography}
\end{document}